# Structure, Magnetism and First Principles Modeling of the Na$_{0.5}$La$_{0.5}$RuO$_3$ Perovskite

Loi T. Nguyen[1], Matthieu Saubanère[2], and Robert J. Cava[1*].

[1]Department of Chemistry, Princeton University, Princeton, New Jersey 08544, USA

[2]Institut Charles Gerhardt, CNRS/Université de Montpellier, Place Eugène Bataillon, F-34095 Montpellier, France

*E-mail : rcava@princeton.edu

**ABSTRACT:** High purity polycrystalline Na$_{0.5}$La$_{0.5}$RuO$_3$ was synthesized by a solid state method, and its properties were studied by magnetic susceptibility, heat capacity and resistivity measurements. We find it to be a tetragonal perovskite, in contrast to an earlier report, with random La/Na mixing. With a Curie-Weiss temperature of -231 K and effective moment of 2.74 µB/mol-Ru, there is no magnetic ordering down to 1.8 K. A broad hump at 1.4 K in the heat capacity, however, indicates the presence of a glassy magnetic transition, which we attribute to the influence of the random distribution of Na and La on the perovskite A sites. Comparison to CaRuO$_3$, a structurally ordered ruthenate perovskite with similar properties, is presented. First-principle calculations indicate that the Na-La distribution determines the local magnetic exchange interactions between Ru ions, favoring either antiferromagnetic or ferromagnetic coupling when the local environment is Na or La rich. Thus our data and analysis suggest that mixing cations with different charges and sizes on the A site in this perovskite results in magnetic frustration through a balance of local magnetic exchange interactions.

## 1. Introduction

ABO$_3$ oxide perovskites have been known and studied for many decades due to their flexible structures and interesting magnetic and electronic properties[1-3]. Specifically, ruthenate perovskites can display unusual electronic and magnetic properties due to the interplay between spin, charge and orbital degrees of freedom and the moderate spin-orbit coupling displayed by the 4$d$ electron based transition metal Ru [4-15]. These properties range from superconductivity in Sr$_2$RuO$_4$[16,17], to ferromagnetism in SrRuO$_3$ (Tc = 165 K) to paramagnetism in CaRuO$_3$ [12,18-21]. The latter two examples show how similar divalent cations (Sr$^{2+}$ and Ca$^{2+}$) in the same crystal structure can give different magnetic properties. The difference in the resulting tilt angles between the rigid RuO$_6$ octahedra due to the accommodation of different size Sr and Ca ions in the perovskite cavities is widely believed to be the structural feature that distinguishes ferromagnetic SrRuO$_3$ from paramagnetic CaRuO$_3$[12,22,23]. Finally, the different polymorphs of BaRuO$_3$ are known to display different electronic and magnetic properties[24-27]. Thus, ruthenate perovskites appear to be an ideal system to study structure-magnetism correlations strongly correlated electrons.

Na$_{0.5}$La$_{0.5}$RuO$_3$ was previously reported to adopt a cubic structure, space group *Pm-3m* with lattice parameter a = b = c = 3.8874 Å. In primitive cubic ABO$_3$ oxide perovskites with unit cell dimensions near 4 angstroms, when two A site ions are present they are randomly mixed on one site and the crystal structure is based on a three-dimensional network of symmetrically equivalent, undistorted BO$_6$ octahedra sharing corners oxygens; between the BO$_6$ octahedra all the B-O-B angles are 180°. Thus when mixtures of A site ions are present in primitive cubic perovskites with unit cell dimensions near 4 angstroms, the mixture must display no long range chemical order, and, further, the BO$_6$ octahedra are neither dimensionally distorted nor tilted [1-3]. This would be the case for Na$_{0.5}$La$_{0.5}$RuO$_3$ in the reported cubic crystal structure. In addition to the crystal structure, the magnetic properties of cubic Na$_{0.5}$La$_{0.5}$RuO$_3$, measured down to 2 K, have previously been reported [28].

Here we synthesize a high purity Na$_{0.5}$La$_{0.5}$RuO$_3$ powder sample through solid state reaction, re-investigate the crystal structure, and perform more detailed material characterization measurements down to 0.37 K. We observe a glassy magnetic transition at around 1.4 K, which we attribute to the influence of the disordered Na-La distribution. First-Principles Density-Functional Theory (DFT) calculations support the presence of a short range correlation between Na and La atoms on the A site, and configurations where each Na is surrounded on average by 3 La and vice versa are found to be the most stable. The calculations show that the Na/La statistical distribution strongly influences the magnetic interactions between the Ru's - locally favoring antiferromagnetic coupling when the local environment is Na rich and ferromagnetic coupling when the local environment is La rich. Generally speaking, the calculations show that system cannot attain an optimal magnetic ordering at low temperature due to the random Na/La distribution, which leads to a competition between local antiferromagnetic and ferromagnetic coupling, magnetic frustration, and a low temperature glassy magnetic state.

## 2. Experiment

The polycrystalline sample of Na$_{0.5}$La$_{0.5}$RuO$_3$ was synthesized by solid-state reaction using Na$_2$CO$_3$ (dried in an oven at 120°C for 3 days before use), dry La$_2$O$_3$, and RuO$_2$ (Alfa Aesar, 99.5%, 99.99%, and 99.95%, respectively) in stoichiometric ratios as starting materials. Reagents were mixed thoroughly, placed in alumina crucibles, and heated in air at 900°C for 24 hours. The resulting powder was reground and heated in air at 1000°C for 24 hours. The phase purity and crystal structure were determined through powder x-ray diffraction (PXRD) using a Bruker D8 Advance Eco with Cu Kα radiation and a LynxEye-XE detector. The crystal structure refinement based on the powder diffraction data was performed with GSAS[29]. The crystal structure drawings were created by using the program VESTA[30].

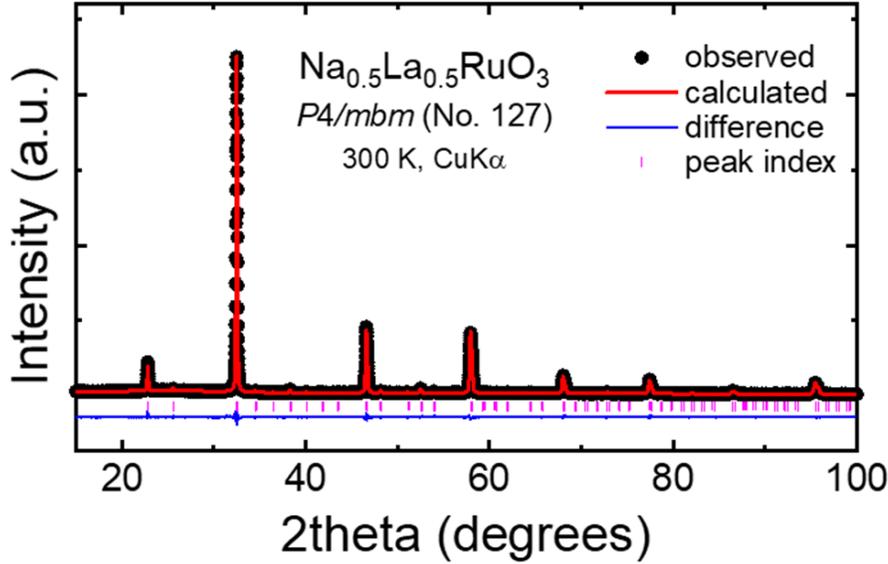

*Figure 1. Rietveld powder x-ray diffraction refinement of the structure of $Na_{0.5}La_{0.5}RuO_3$ in space group P4/mbm: $\chi^2$ = 2.15, $R_{wp}$ = 5.32%, $R_p$ = 3.94%, $R_F^2$ = 3.54%.*

The magnetic susceptibility of $Na_{0.5}La_{0.5}RuO_3$ was measured in a Quantum Design Physical Property Measurement System (PPMS) DynaCool equipped with a VSM option. The magnetic susceptibility between 1.8 and 300 K, defined as M/H, where M is the sample magnetization and H is the applied field, was measured at different applied magnetic fields. The sample was pressed, sintered, and cut into pieces with the approximate size 1.4 × 2.5 × 1.0 mm³ for the resistivity measurements. Four Pt contact wires were connected to the samples using silver paint. The resistivity was measured by the dc four-contact method in the temperature range 1.8 to 300 K in the PPMS. The specific heat was measured from 200 to 1.8 K by the PPMS DynaCool equipped with a heat-capacity option, down to 0.37 K by using a ³He system. Similar sample preparation and measurements were used to facilitate comparison to $CaRuO_3$.

### 3. Results and discussion

### 3.1 Crystal structure

$Na_{0.5}La_{0.5}RuO_3$ was previously reported to adopt a small unit cell cubic perovskite structure, space group $Pm\text{-}3m$ with a lattice parameter of 3.887 Å[28]. This is the basic perovskite subcell, but the presence of many weak X-ray peaks in our powder diffraction pattern indicates that our material is not cubic. **Figure 1** shows the structural refinement of $Na_{0.5}La_{0.5}RuO_3$ obtained by using a tetragonal perovskite cell, with $a = b \approx \sqrt{2}a_{cubic}$, and $c \approx 2c_{cubic}$ in space group $P4/mbm$. Although a small cubic symmetry cell matches the positions of the stronger reflections, **Figure 2** clearly shows the better fit of the primitive tetragonal $P4/mbm$ cell to the diffraction data, which, in addition to the strong reflections, indexes all the minor reflections. The peaks are not matched by a body centered tetragonal cell, common for tetragonal

*Table 1. Structural parameters for $Na_{0.5}La_{0.5}RuO_3$ at 300 K. Space group P4/mbm (No. 127).*

| Atom | Wyckoff. | Occ. | x | y | z | $U_{iso}$ |
|---|---|---|---|---|---|---|
| Na/La | 4f | 0.5/0.5 | ½ | 0 | 0.24243(3) | 0.0236(3) |
| Ru1 | 2a | 1 | 0 | 0 | 0 | 0.0455(7) |
| Ru2 | 2b | 1 | 0 | 0 | ½ | 0.0108(5) |
| O1 | 4e | 1 | 0 | 0 | 0.2522(2) | 0.0273(5) |
| O2 | 4h | 1 | 0.3709(1) | 0.8709(1) | ½ | 0.0309(3) |
| O3 | 4g | 1 | 0.2144(2) | 0.7144(2) | 0 | 0.0757(5) |

a = b = 5.5071(1) Å, c = 7.7830(4) Å, V = 236.040(2) Å³, α = β = γ = 90°

$\chi^2$ = 2.15, $R_{wp}$ = 5.32%, $R_p$ = 3.94%, $R_F^2$ = 3.54%

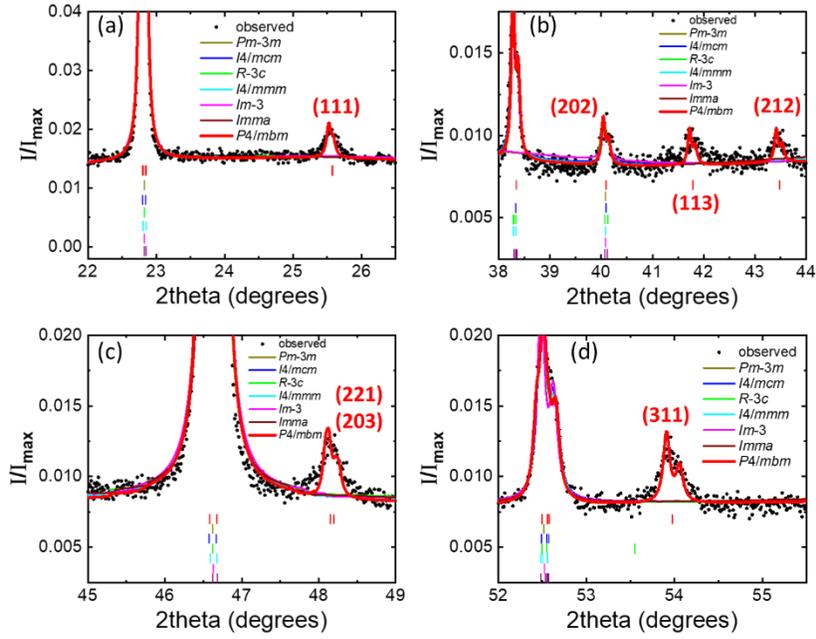

**Figure 2.** Le Bail fits of our diffraction data for $Na_{0.5}La_{0.5}RuO_3$ in space group $Pm\text{-}3m$ and its six subgroups. The space group $Pnma$ can also index the peaks (not shown) indexed by $P4/mbm$, but more reflections will be present due to its lower symmetry; there is no indication of dimensional orthorhombicity. (see Supplemental Information, **Figure S2**).

perovskites, as peaks that clearly violate the body centering conditions are observed.

Finally, because it is a common space group for distorted perovskites, we tested the agreement of a structural model in the orthorhombic space group $Pnma$ with the diffraction data (**Figure S1**). The difference between the $Pnma$ and $P4/mbm$ structural models has to do with a difference in the position of one of the oxygen atoms, which is beyond the sensitivity of our experiments to resolve, but because the material is clearly dimensionally tetragonal and because the position obtained for one of the oxygen atoms in our $Pnma$ refinement doesn't make sense due to the existence of an unphysical O-Ru-O angle, we employ the tetragonal model as our final structure. Comparison of Le Bail fittings of the diffraction data for $Na_{0.5}La_{0.5}RuO_3$ with space group $Pm\text{-}3m$ and its six subgroups supports the space group $P4/mbm$ as being the best space group for refining the structure. The refinement and structural parameters of $Na_{0.5}La_{0.5}RuO_3$ based on the tetragonal model are summarized in **Table 1**. The interatomic distances of Ru-O and Na/La-O in $Na_{0.5}La_{0.5}RuO_3$ at 300 K, which all make sense, are outlined in **Table 2**. The analogous mixed Na-La titanate perovskite $Na_{0.5}La_{0.5}TiO_3$ [31] has similarly been reported to display non-cubic symmetry. That material is rhombohedral however, which is not the case for the current material. Why the low intensity diffraction peaks that clearly require a non-cubic larger volume unit cell for our $Na_{0.5}La_{0.5}RuO_3$ were not considered in the previous report is not clear, but may be attributed to differences in the synthetic conditions. The refined structural and atomic separation parameters for space group *Pnma* are shown in the SI, **Tables S1 and S2**.

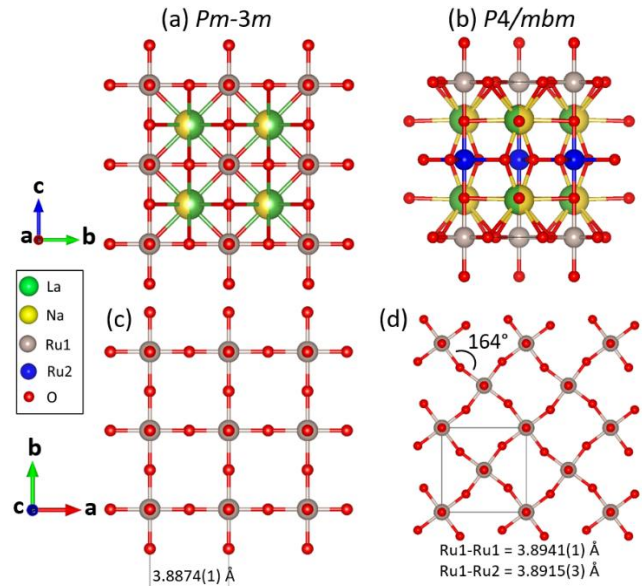

**Figure 3.** (Top) The crystal structure of $Na_{0.5}La_{0.5}RuO_3$ based on the reported cubic space group $Pm\text{-}3m$ and tetragonal symmetry structure displayed by our material (space group $P4/mbm$). (Bottom) Projections down the c-axis show no octahedral tilting in the $Pm\text{-}3m$ space group and the in-plane tilting of the $RuO_6$ octahedra in the $P4/mbm$ model.

**Table 2.** Selected interatomic distances (Å) in space group $P4/mbm$ in $Na_{0.5}La_{0.5}RuO_3$ at 300 K.

| Interatomic distance (Å) | |
|---|---|
| Ru1-O1 (x2) | 1.96(1) |
| Ru1-O3 (x4) | 1.97(9) |
| Ru2-O1 (x2) | 1.93(1) |
| Ru2-O2 (x4) | 2.16(8) |
| Na/La-O1 (x4) | 2.75(4) |
| Na/La-O2 (x2) | 2.24(4) |
| Na/La-O3 (x2) | 2.52(6) |

The crystal structures of cubic and tetragonal $Na_{0.5}La_{0.5}RuO_3$ are compared in **Figures 3a-d**. The differences primarily arise from the displacement of some of the oxygens in the tetragonal model and the tilt of $RuO_6$ octahedra from their ideal cubic orientations. The $RuO_6$ octahedral tilting in this tetragonal symmetry structure is only around one crystallographic axis. This can be compared to the multiple axis tilting found in orthorhombic $CaRuO_3$ and $SrRuO_3$, and the absence of any tilting in the small-cell primitive cubic perovskite. The in-phase rotation of the $Ru(1)O_6$ octahedra around the c axis makes this a $a^0a^0c^+$ system[32-35] in the standard perovskite nomenclature. The Ru(1)-O-Ru(1) bond angle is calculated to be 164° as displayed in **Figure 3d**. The slight compression of the $RuO_6$ octahedra expected for $Ru^{4+}$ in a high spin configuration, which leads to slightly shorter axial than equatorial Ru-O bond lengths is observed and summarized in **Table 2**. Because oxygen is a relatively weak scatterer of X-rays compared to Ru and La, determination of the oxygen positions to higher precision will require diffraction data that is beyond the scope of the present study.

### 3.2 Magnetic and electronic properties

**Figure 4a** shows the temperature-dependent magnetic susceptibility of $Na_{0.5}La_{0.5}RuO_3$ from 300 to 1.8 K, measured under the applied field of 1 kOe. There is no magnetic ordering down to 1.8 K. The inset in **Figure 4a** displays the ZFC/FC susceptibility from 50-1.8 K, under an applied magnetic field of 100 Oe. The absence of a bifurcation indicates there is no glassy state present above 1.8 K. **Figure 4b** shows the field-dependent magnetization of $Na_{0.5}La_{0.5}RuO_3$ at various temperatures. All the curves are linear, implying that $Na_{0.5}La_{0.5}RuO_3$ remains paramagnetic down to 2 K. Resistivity measurements on $Na_{0.5}La_{0.5}RuO_3$ from 300 to 1.8 K are shown in **Figure 4c**. The data in the elevated temperature regime (250–300 K) fit to an activated model where $\rho = \rho_0 \times e^{\frac{E_a}{k_B T}}$ (red line) with $E_a$ = 1.38 meV. In comparison $Li_2RuO_3$ has an activation energy of 53 meV [36]. The small resistivity gap of 1.38 meV suggests that $Na_{0.5}La_{0.5}RuO_3$ is near the boundary of metal and semiconductor. The normalized temperature dependence of the resistance of a similarly prepared $CaRuO_3$ sample exhibits similar R/R(300K) value from 300 K down to 100 K (not shown). Below that, $Na_{0.5}La_{0.5}RuO_3$ shows a slightly larger resistance than $CaRuO_3$. As usual, measurements on single crystals are required to determine the absolute temperature-dependent resistivity of the system without the influence of grain boundary scattering.

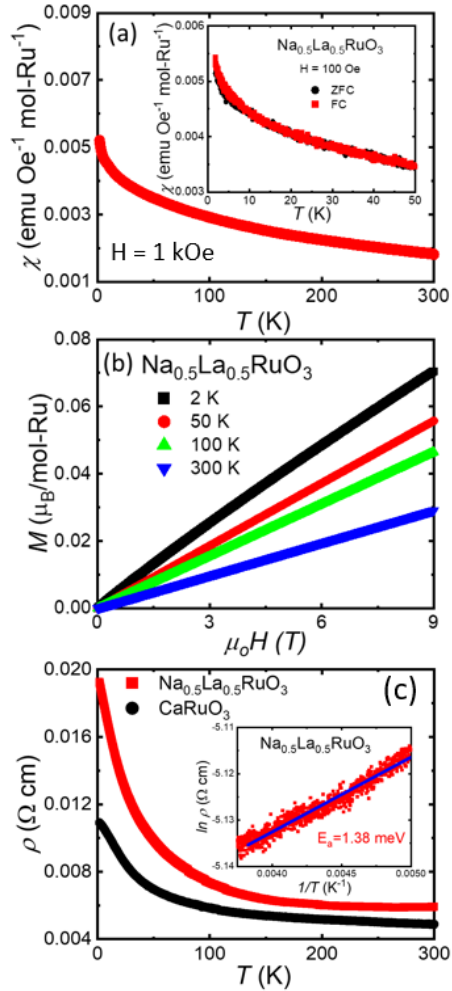

*Figure 4.* (a) Temperature-dependent magnetic susceptibility of $Na_{0.5}La_{0.5}RuO_3$ measured under the applied field of 1 kOe. The inset shows the ZFC/FC susceptibility in an applied field of 100 Oe. (b) Magnetization of $Na_{0.5}La_{0.5}RuO_3$ as the function of the applied field from 0 to 9 T, at various temperatures. (c) Resistivity measurement on polycrystalline $Na_{0.5}La_{0.5}RuO_3$ from 1.8 to 300 K; measurements on a $CaRuO_3$ sample prepared the same way are included for comparison.

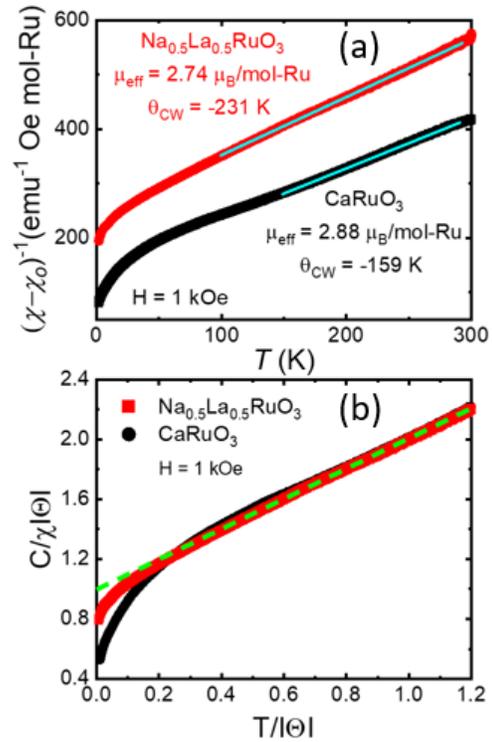

*Figure 5.* Comparison of the magnetic properties of $Na_{0.5}La_{0.5}RuO_3$ and $CaRuO_3$. (a) Inverse magnetic susceptibility, (b) Normalization and comparison of the temperature dependent susceptibilities of these two materials. (The dashed line represents ideal Curie-Weiss behavior.)

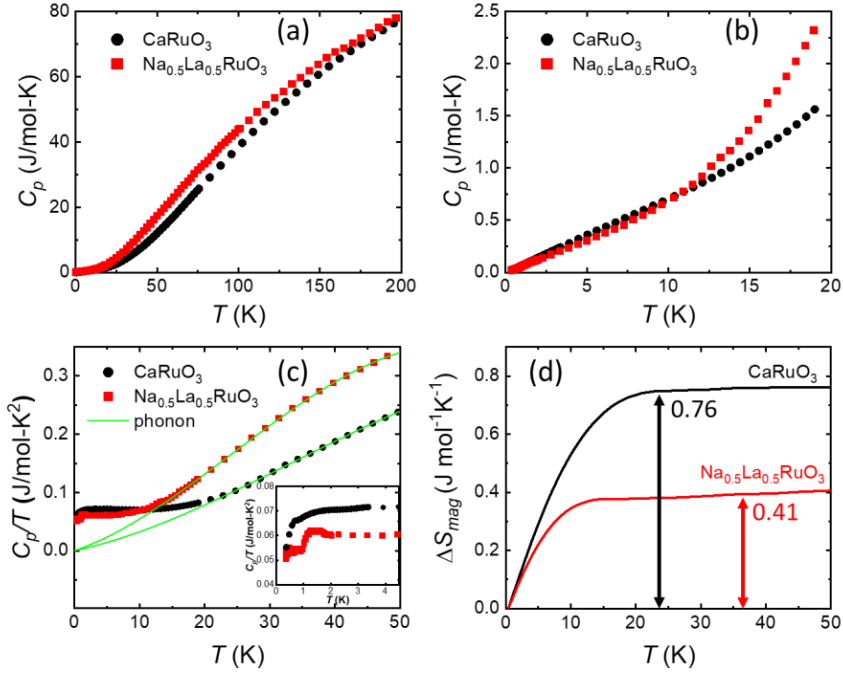

*Figure 6. (a) Molar heat capacity of $Na_{0.5}La_{0.5}RuO_3$ and $CaRuO_3$ between 0.37 and 200 K. (b) Heat capacity below 20 K. (c) Heat capacity by divided temperature, with the phonon fits (green lines). (d) Calculated magnetic entropy for both $Na_{0.5}La_{0.5}RuO_3$ and $CaRuO_3$ up to 50 K. For a two state system the expected integrated entropy is 5.76 J $mol^{-1}K^{-1}$. Both materials display metallic conductivity so there is also likely a temperature independent term present at low temperatures.*

**Figure 5a** demonstrates the magnetic comparison between $Na_{0.5}La_{0.5}RuO_3$ and $CaRuO_3$. The Curie-Weiss fitting produces an effective moment of 2.74 μB/mol-Ru and the Curie-Weiss temperature of -231 K in $Na_{0.5}La_{0.5}RuO_3$ while they are 2.88 μB/mol-Ru and -159 K in $CaRuO_3$, respectively. These values agree well with previous reports [28]. Both materials have a very large negative Curie-Weiss temperatures, indicating strong dominantly antiferromagnetic interactions. However, the absence of magnetic ordering down to 1.8 K implies that the magnetic interactions are strongly frustrated (the frustration index, $f = |\Theta_{CW}|/T_M$, is larger than 127 for $Na_{0.5}La_{0.5}RuO_3$ and 88 in $CaRuO_3$). Comparing to other $Ru^{4+}$ compounds, $Li_2RuO_3$ for example has an effective moment of 2.17 μB/mol-Ru and there is no magnetic ordering down to 5 K [36]. The origin of the temperature-dependent magnetic susceptibility in $CaRuO_3$ has been the matter of research for many years. $Na_{0.5}La_{0.5}RuO_3$ is quite similar although it has the added complexity of the Na/La structural disorder.

**Figure 5b** shows the normalized magnetic susceptibility ($C/\chi|\Theta| = T/|\Theta|+1$) for both $Na_{0.5}La_{0.5}RuO_3$ and $CaRuO_3$. Ideal Curie-Weiss behavior will be a straight line with the slope of 1 and the y-intercept of 1 or -1 in the case of dominant AFM or FM interactions, respectively, on such a normalized plot. As seen in **Figure 5b**, for values of $T/|\Theta_{CW}|$ larger than 0.7, both $Na_{0.5}La_{0.5}RuO_3$ and $CaRuO_3$ follow ideal Curie-Weiss paramagnetism behavior (green dashed line). While $Na_{0.5}La_{0.5}RuO_3$ remains paramagnetic down to a $T/|\Theta_{CW}|$ of 0.3, a small positive deviation of $T/|\Theta_{CW}|$ in $CaRuO_3$ for $T/|\Theta_{CW}|$ between 0.3 and 0.7 indicates the presence of antiferromagnetic short-range correlations in addition to those expected for an ordinary antiferromagnet. Below the $T/|\Theta_{CW}|$ value of 0.3, both $Na_{0.5}La_{0.5}RuO_3$ and $CaRuO_3$ show a negative deviation from the ideal line, indicating the presence of increasingly influential ferromagnetic interactions[37-39].

The heat capacities of $Na_{0.5}La_{0.5}RuO_3$ and $CaRuO_3$ between 1.8 and 200 K are plotted in **Figure 6a**. **Figure 6b** shows an expansion of the heat capacity data below 20 K. At high temperatures, due to the stronger phonon vibrations (larger molar mass), $Na_{0.5}La_{0.5}RuO_3$ has a larger heat capacity than $CaRuO_3$, but they become comparable at around 10 K. As seen in the inset of **Figure 6c**, a broad hump is observed in both compounds, but it is shifted to the lower temperature of around 1.4 K in $Na_{0.5}La_{0.5}RuO_3$. This implies a glassy transition state in $Na_{0.5}La_{0.5}RuO_3$. We attribute this to the Na/La disorder, due to insights gained in our DFT calculations. A fifth-order polynomial is used to fit to the total heat capacity between 50 and 200 K for both compounds. The coefficients, summarized in **Table S3**, are used for extrapolation below 50 K, allowing for an approximate subtraction of the phonon contribution to the low temperature heat capacity data. The green lines in **Figure 6c** account for the phonon contributions. By integrating the $C_{mag}/T = C_{total}/T - C_{phonon}/T$ up to 50 K, the magnetic entropy, outlined in **Figure 6d** for both $Na_{0.5}La_{0.5}RuO_3$ and $CaRuO_3$, recovers to 0.41 J/mol-K and 0.76 J/mol-K, respectively. These calculated values for the magnetic entropy are much smaller than the expected value for a localized spin-1 system. Why these values are so small is not known - this may be an interesting question for later theoretical investigation of the ruthenate family to address. Fitting the heat capacity data with $C_p/T = \gamma + \beta T^2$ yields the Sommerfeld constant γ of 39 mJ mol$^{-1}$K$^{-2}$ for $Na_{0.5}La_{0.5}RuO_3$ and 61 mJ mol$^{-1}$K$^{-2}$ for $CaRuO_3$. While the former value is first reported, the latter one agrees with previous studies[40,41].

### 3.3 Theoretical calculations

Calculations were performed to address the thermodynamics of $Na_{1/2}La_{1/2}RuO_3$, specifically to address the origin of the random La-Na distribution experimentally found in the material. In the $ABO_3$ Perovskite structure, each A atom is surrounded in by 6 other A's in the first neighbor shell. In order to obtain a representative statistical sample of the different local Na/La distributions possible in the material, we constructed large cubic super-cells containing 64 atoms of Ru, where 32 Na and 32 La are randomly distributed. Then all the structures were fully relaxed using the PBE exchange and correlation functional [42–44] with a Hubbard correction $U_{eff}$ = 2.25 eV to account for the electron-electron repulsion in the Ru 4$d$ orbitals [45]. In order to estimate the formation energy independently from the magnetic ordering, we compared the energy of the non-magnetic phases with respect to the decomposition reaction:

$Na_{1/2}La_{1/2}RuO_3 \rightarrow$ ¼ $(Na_2O + La_2O_3) + RuO_2$   (1)

The results are shown in **Figure 7a**, and demonstrate that long-range Na-La interactions are rather weak, since the formation energy seems to depend quasi-linearly on $N_k$ in two regimes, $N_k$ being the average number of the same type of nearest neighbor per Na/La. In the homogeneously charged regime ($0 \leq N_k \leq 3$) the charge around the Ru is homogeneously distributed, each Ru being 4+. This is in contrast to the segregated regime ($3 < N_k \leq 6$) where an inhomogeneous charge distribution is observed. The local Na/La ordering results from the compromise between the two regimes in order to equalize homogeneously the charges over all Ru, and to minimize the local mechanical stress due to the difference of size and charge between Na and La. From a combinatory point of view, among the $N_b$ = $3N$ Nearest Neighbor (NN) bonds available in the Perovskite cubic cell containing $N$ sites statistically occupied by Na and La, one can distribute $N_{Na-La}$ Na-La bonds, $0$ $(N_k \vec{N} \rightarrow \infty 0) < N_{Na-La} \leq N_b(N_k=6)$. Then, the largest configuration space is defined for $N_{Na-La} = N_b/2$ ($N_k$=3), and contains binomial ($N_b$, $N_b/2$) configurations. As seen in **Figure 7a**, the small energy difference between the configurations close to or having $N_k$=3 should lead to an important contribution of configurational entropy.

We next consider the magnetic behavior of $Na_{1/2}La_{1/2}RuO_3$, which shows dominant antiferromagnetic interactions and strong magnetic frustration at low temperature. In comparison, recall that the Ru-based perovskites $CaRuO_3$ and $SrRuO_3$ show different magnetic behavior, the former being paramagnetic at low temperature with dominant antiferromagnetic interactions, while the latter is ferromagnetic. To understand such different behavior, we first compare their crystallographic structures and observe that the mean distance Ru-O remains equivalent for all compounds (~1.95 to 1.99 Å) suggesting equivalent Ru-O covalency. Note that due to the partial filling of the Ru $t_{2g}$ orbitals (Ru$^{4+}d^4$), a $D_{4h}$ Jahn-Teller (JT) distortion occurs and leads to a compression of the $RuO_6$ octahedra, which favors antiferromagnetic over ferromagnetic exchange coupling. The variation of the Ru-O-Ru angle should then drastically affect the magnetic ordering, since it might balance the competition between indirect exchange interactions and direct exchange interactions. Indeed, AFM contributions to the indirect exchange interaction are maximal for $Ru - \hat{O} - Ru$ = 180° while minimal for $Ru - \hat{O} - Ru$ = 90°. Note that the occurrence of the JT $D_{4h}$ distortion also affects the indirect exchange correlation, favoring AFM over FM interactions due to orbital selectivity, the doubly occupied $t_{2g}$ orbital being orthogonal to the JT compression axis (**Figure S3**). The direct exchange interaction is related to the direct hopping between Ruthenium and stabilizes the anti-ferromagnetic ordering. It is characterized by energy of about $\frac{t_{dd}^2}{U}$ where $U$ is the Coulomb electron-electron repulsion, and $t_{dd}$ refers to the $d$ orbital overlap between NN Ru. $t_{dd}$ increases when closing the Ru-O-Ru angle since Ru-Ru distances are concomitantly decreased.

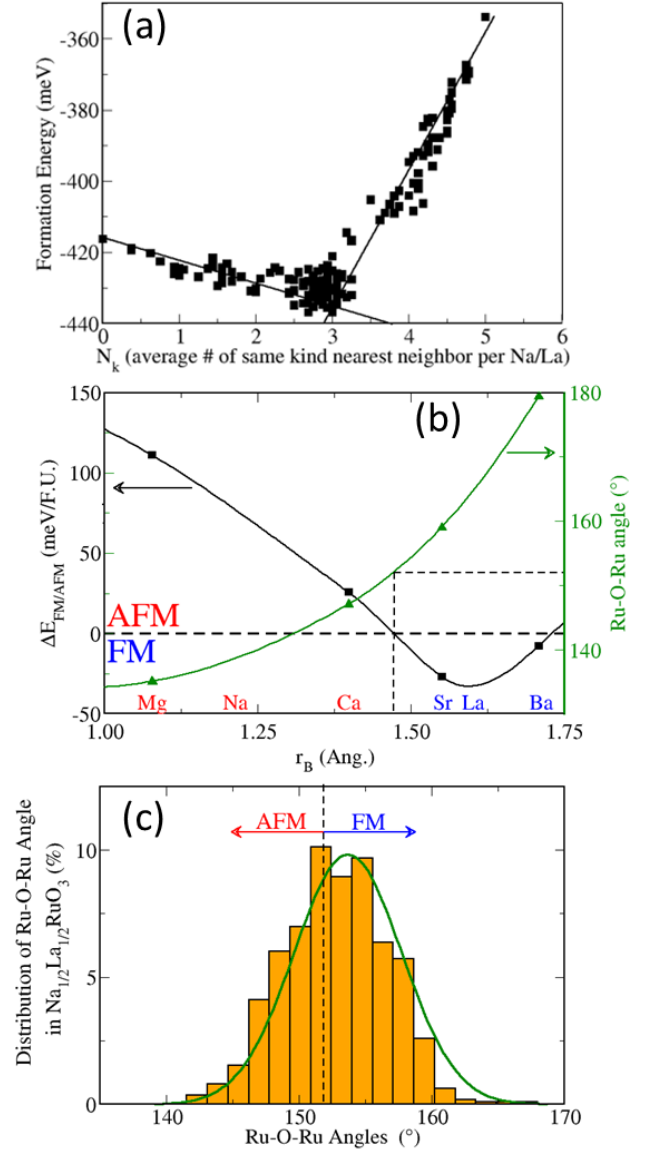

**Figure 7**. (a) Formation energy in meV as a function of $N_k$, the average of the same type of nearest neighbor, from DFT + U calculations. The minimum energy is found where each Na (La) is surrounded on average by 3 La (3 Na). (b) Magnetic ordering energy (black square, left axis) and Ru-O-Ru angle (green triangle, right axis) as the function of the Bader radii for $ARuO_3$ (A=Mg, Ca, Sr and Ba). Lines are guide for the eyes. (c) Histogram of the statistical distribution of Ru-O-Ru angles in the five most stable calculated $Na_{1/2}La_{1/2}RuO_3$ structures having $N_k$=3. The green line shows the Gaussian fit of the data and the dashed black line shows the FM/AFM transition observed in $ARuO_3$ (A=Mg, Ca, Sr and Ba).

In order to investigate the magnetic ordering in Ru-based perovskites, we show in **Figure 7b**, $\Delta E_{FM/AFM}$, the energy difference between the ferromagnetic and the Néèl anti-ferromagnetic orders for the series $ARuO_3$ (A = Mg, Ca, Sr and Ba) as computed with DFT+U using $U_{eff}$ = 2.25 eV on Ru $d$ orbitals. The $U_{eff}$ value used has been carefully chosen in order to agree with measured band-gap and magnetic order determined in $CaRuO_3$ and $SrRuO_3$ (**Figures S4 and S5**). In $BaRuO_3$ ferromagnetism is shown to be favored, with a $Ru - \hat{O} - Ru$ = 180°. Substituting Ba by Sr results in a decrease of the Ru-O-Ru angle and a concomitant decrease of $\Delta E_{FM/AFM}$ due to the loss of Ruthenium Oxygen orbital overlap, reducing the AFM contributions to the indirect exchange interaction. Closing further the Ru-O-Ru angle by substituting Sr by Ca decreases sufficiently the Ru-Ru distances so that the contribution of the AFM direct exchange interaction becomes significant. Consequently, $\Delta E_{FM/AFM}$ increases, and the AFM state is favored over the FM state for $Ru - \hat{O} - Ru$ <152°. A transferable relation between the Ru-O-Ru angle and the size of the alkali-earth A is obtained by looking the volume of its Bader basin $V_B$ and the associated Bader ionic radius [46] $r_B = \sqrt[3]{\frac{3}{4} V_B}$. As shown in **Figure S6**, the Bader ionic radii of the considered Alkali-earth are linearly correlated with their standard ionic radii. Our results suggest that the FM/AFM transition occurs for $r_B$ =1.47Å, such that Ru-based perovskites with smaller alkali-earth ($Mg^{2+}$, $Ca^{2+}$) show AFM ordering while Ru-based perovskites with bigger alkali-earth ($Sr^{2+}$, $Ba^{2+}$) show FM ordering. Note that the low temperature paramagnetism in $CaRuO_3$ might originate from the occurrence of second NN exchange interactions, which are enhanced due to the JT distortion. Indeed second NN interactions might stabilize other AFM configurations with energy close to or even lower than the Néèl configuration.

The magnetic properties of $Na_{1/2}La_{1/2}RuO_3$ can be understood in light of this analysis. Indeed, as shown in **Figure 7b** the effective Bader ionic radius of Na and La in $Na_{1/2}La_{1/2}RuO_3$ are significantly different, being much larger for La (1.56 Å) than for Na (1.20 Å). The result is that the Bader radii of Na (La) is lower (larger) than the one of $Ca^{2+}$ ($Sr^{2+}$), respectively. Consequently, as shown in **Figure 7c** the Ru-O-Ru angles show values distributed between 140° and 165° depending whether the local environment is Na or La excess. In conclusion, the magnetic interactions between Ruthenium with a smaller (larger) Ru-O-Ru angle should locally favor antiferromagnetic (ferromagnetic) coupling when the local environment is Na (La) excess. This leads to a global antiferromagnet for our material, with, however, strong magnetic frustration present because the Ru cannot display optimal magnetic ordering due to the different Na/La local order that occurs in the structure. Our calculations show that mixing cations on the A site with different charge and size in the Ru based perovskites should lead to magnetic frustration by locally controlling the magnetic exchange interactions.

### 4. Conclusion

The crystal structure of high purity $Na_{0.5}La_{0.5}RuO_3$, synthesized at 1000°C in air was investigated and found to adopt the tetragonal *P4/mbm* space group. The electronic and magnetic properties were characterized by magnetic susceptibility, heat capacity and resistivity measurements. With a Curie-Weiss temperature of -231 K and an effective moment of 2.74 µB/mol-Ru, magnetic ordering is not observed down to temperatures of 1.8 K. However, a broad hump at 1.4 K in the heat capacity is present, which we interpret as being due to a glassy magnetic transition due to the random distribution of Na and La on the perovskite A site. Our calculations support the interpretation that Na/La short-range ordering occurs and locally controls the magnetic exchange interactions between Ru ions. This indicates that mixing cations with different charges and sizes results in the magnetic frustration - the system cannot afford an optimal magnetic ordering at low temperature due to the different local Na/La distributions.

## AUTHOR INFORMATION

### Corresponding Author

* rcava@princeton.edu (Robert J. Cava)

### Author Contributions

The manuscript was written through contributions of all authors. All authors have given approval to the final version of the manuscript.

### Funding Sources

The US Department of energy, grant number DE FG02 98ER 45706.

## ACKNOWLEDGMENT

This research was supported by the Division of basic energy sciences of the US Department of energy, grant number DE FG02 98ER 45706.

Table of Contents artwork

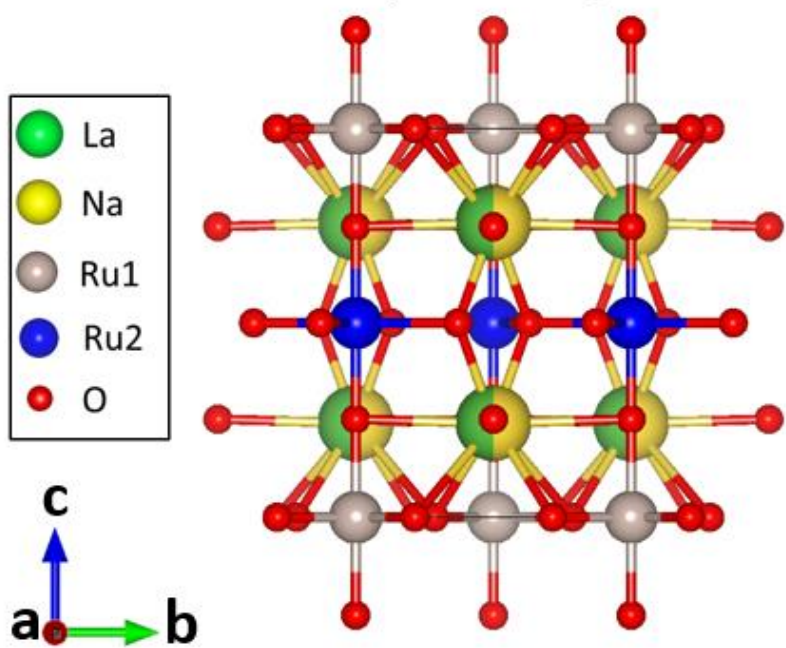

# Supplemental Information

**Structure, Magnetism and First Principles Modeling of the Na$_{0.5}$La$_{0.5}$RuO$_3$ Perovskite**

Loi T. Nguyen[1], Matthieu Saubanère[2], and Robert J. Cava[1*].

[1]Department of Chemistry, Princeton University, Princeton, New Jersey 08544, USA

[2]Institut Charles Gerhardt, CNRS/Université de Montpellier, Place Eugène Bataillon, F-34095 Montpellier, France

*E-mail : rcava@princeton.edu

**Table S1**: Structural parameters for Na$_{0.5}$La$_{0.5}$RuO$_3$ at 300 K in space group *Pnma* (No. 62).

| Atom  | Wyckoff. | Occ.    | $x$         | $y$      | $z$        | $U_{iso}$ |
|-------|----------|---------|-------------|----------|------------|-----------|
| Na/La | 4c       | 0.5/0.5 | -0.02136(2) | ¼        | -0.0057(6) | 0.0158(3) |
| Ru    | 4b       | 1       | 0           | 0        | ½          | 0.0282(3) |
| O1    | 4c       | 1       | 0.0128(2)   | ¼        | 0.3596(2)  | 0.0331(1) |
| O2    | 8d       | 1       | 0.2773(3)   | -0.0207(1) | 0.2768(2) | 0.0331(1) |

$a = 5.51352(2)$ Å, $b = 7.78580(3)$ Å, $c = 5.50359(1)$ Å, $V = 236.254(1)$ Å$^3$, $\alpha = \beta = \gamma = 90°$
$\chi^2 = 2.01$, $R_{wp} = 5.13\%$, $R_p = 3.89\%$, $R_F^2 = 4.73\%$

**Table S2**: Selected interatomic distances (Å) in space group *Pnma* in Na$_{0.5}$La$_{0.5}$RuO$_3$ at 300 K.

| Interatomic distance (Å) | |
|---|---|
| Ru1-O1 (x2) | 2.10(4) |
| Ru1-O2 (x4) | 1.96(1) |
| Na/La-O1 (x2) | 2.02(1) |
|  | 2.69(1) |
| Na/La-O2 (x6) | 2.54(1) |
|  | 2.70(1) |
|  | 2.72(1) |

**Table S3**: Output parameters from the fifth-order polynomial fittings to the heat capacity of $Na_{0.5}La_{0.5}RuO_3$ and $CaRuO_3$ from 50-200 K. ($C_p/T = a_0 + a_1*T + a_2*T^2 + a_3*T^3 + a_4*T^4 + a_5*T^5$)

|       | $Na_{0.5}La_{0.5}RuO_3$ | $CaRuO_3$ |
|-------|-------------------------|-----------|
| $a_0$ | -0.205                  | -0.215    |
| $a_1$ | 0.021                   | 0.014     |
| $a_2$ | -2.594E-4               | -1.247E-4 |
| $a_3$ | 1.707E-6                | 5.599E-7  |
| $a_4$ | -5.852E-9               | -1.335E-9 |
| $a_5$ | 8.191E-12               | 1.354E-12 |

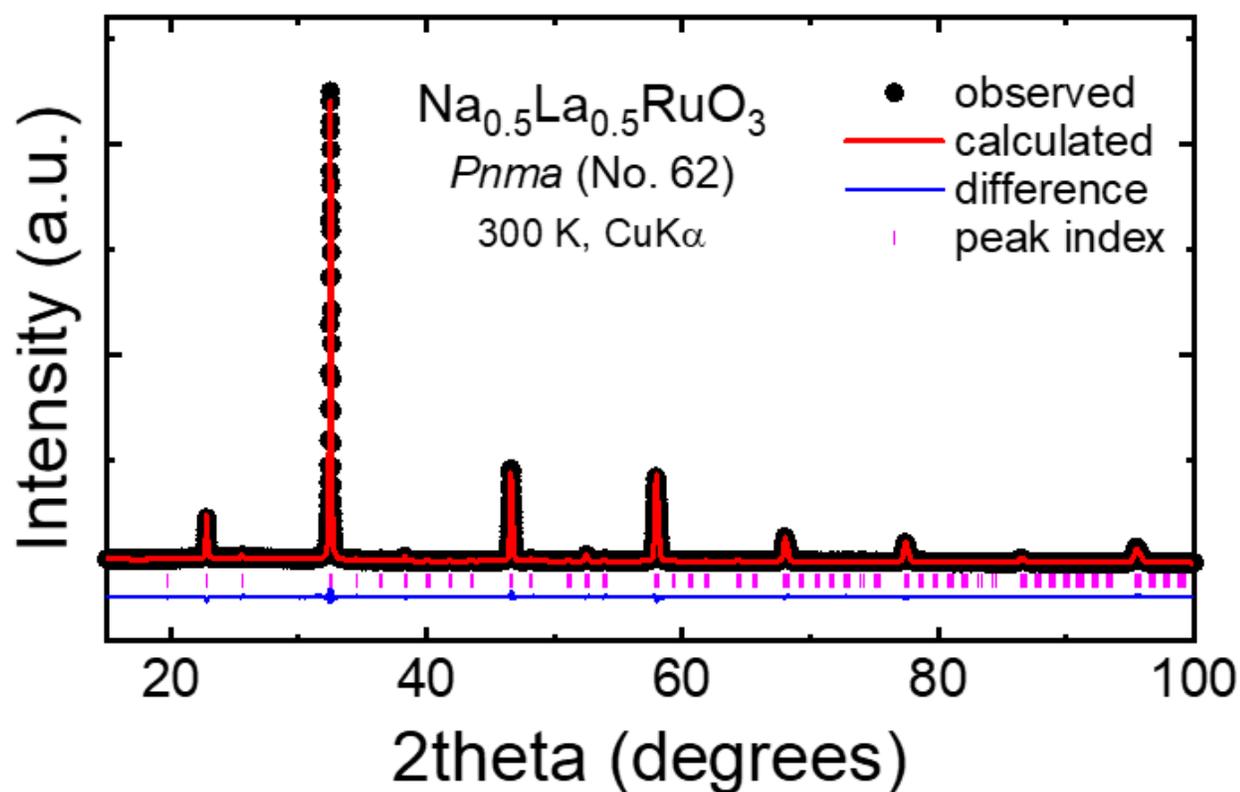

**Figure S1**: Rietveld powder x-ray diffraction refinement of the alternative but less successful model of $Na_{0.5}La_{0.5}RuO_3$ in space group *Pnma*: $\chi^2 = 2.01$, $R_{wp} = 5.13\%$, $R_p = 3.89\%$, $R_F^2 = 4.73\%$.

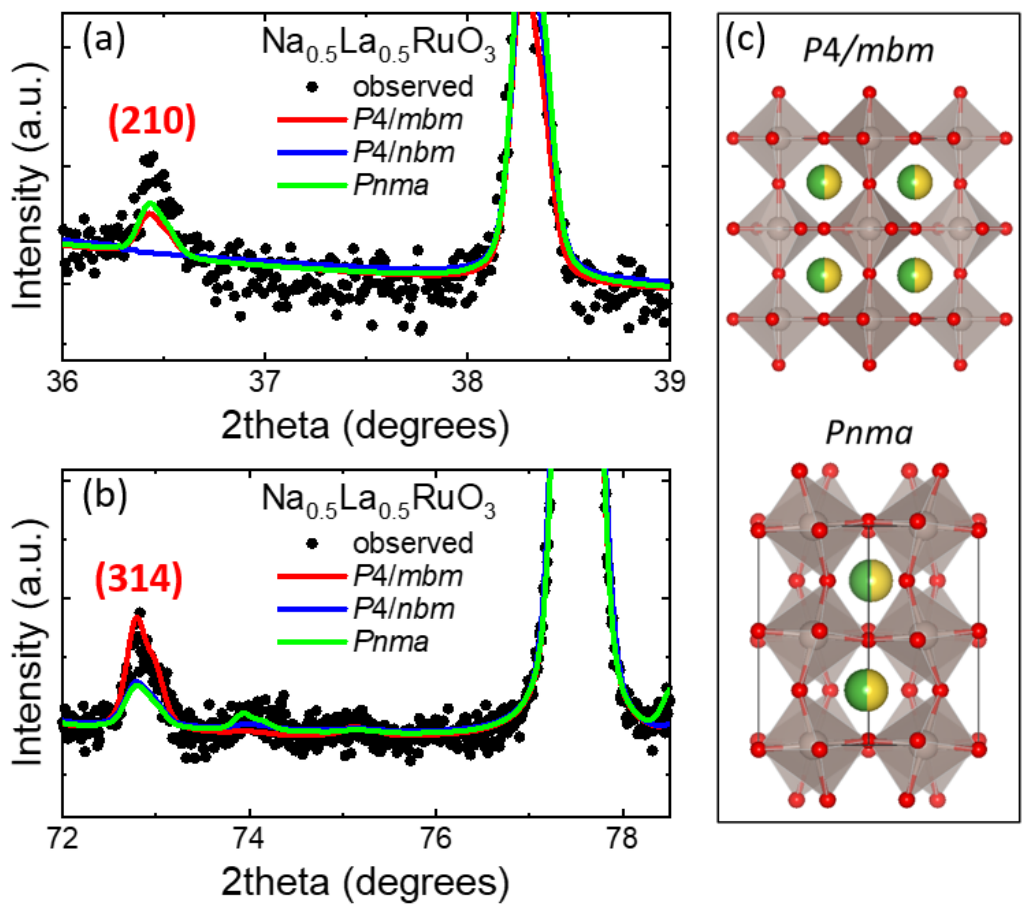

**Figure S2**: (a) and (b) Comparison of the Rietveld fittings for $Na_{0.5}La_{0.5}RuO_3$ in some common subgroups (*P4/mbm*, *P4/nbm* and *Pnma*) of the cubic space group *Pm-3m*. Space group *Pnma* can index all the peaks that *P4/mbm* can, however, there are more calculated reflections present due to its lower symmetry. (c) comparison of the crystal structures of $Na_{0.5}La_{0.5}RuO_3$ in space groups *P4/mbm* and *Pnma*.

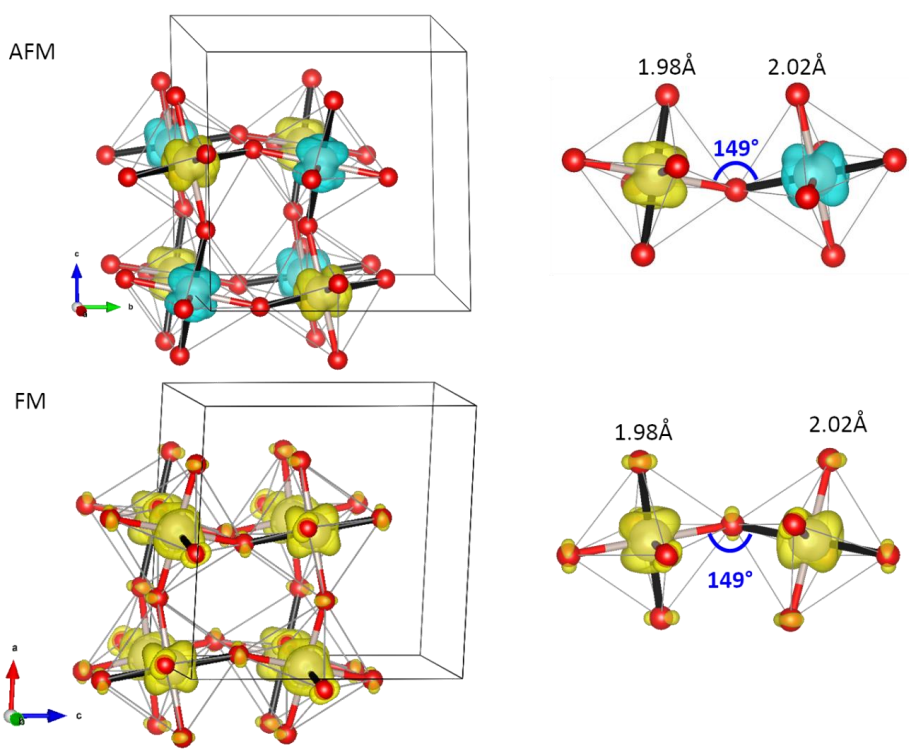

**Figure S3**: Magnetization density plot for the anti-ferromagnetic state (top) and the ferromagnetic state (bottom) of $CaRuO_3$ as computed with DFT+U with $U_{eff}$ = 2.25 eV. Black bonds refer to shorter Ru-O bonds along the JT $D_{4h}$ compression axis. Ru-O distances and Ru-O-Ru angle are highlighted. Similar organization of the JT compressions axis is found for $SrRuO_3$. Note that ferromagnetic states are found metallic without JT distortion for $U_{eff}$ < 2 eV.

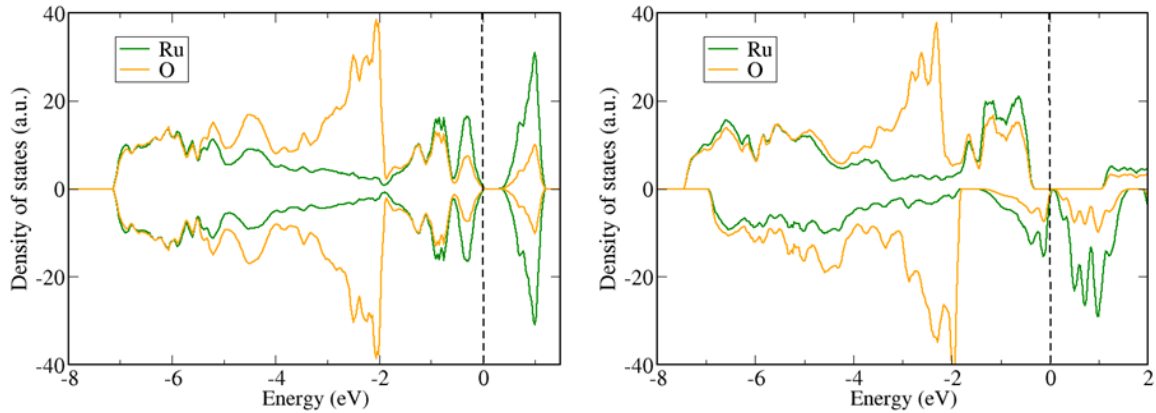

**Figure S4**: Density of states of the anti-ferromagnetic state (left) and the ferromagnetic state (right) of $CaRuO_3$ as computed with DFT + U with $U_{eff}$ = 2.25 eV. The dashed black line highlight the Fermi level. Band gaps are originated from the JT distortion that splits the partially filled $t_{2g}$ states of $Ru^{4+}$.

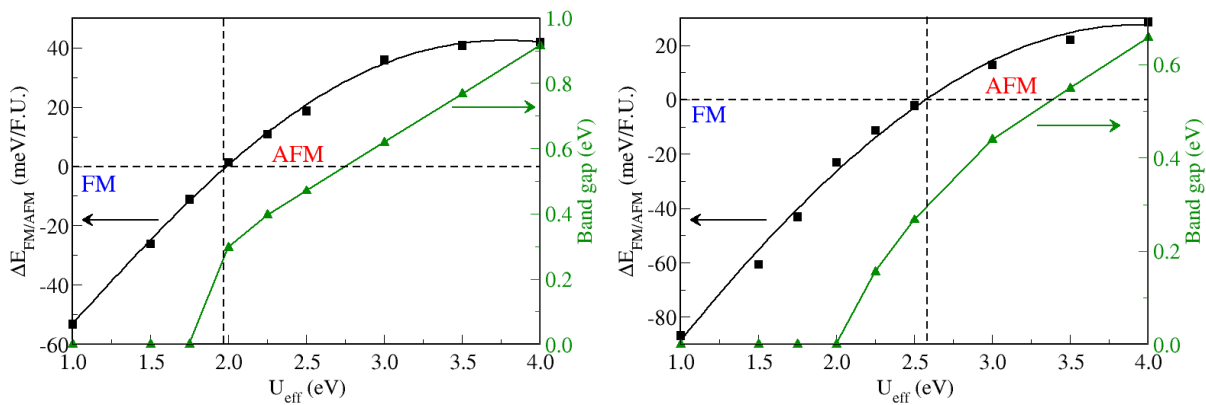

**Figure S5**: Magnetic ordering energy $\Delta E_{FM/AFM} = E_{FM} - E_{AFM}$ (black curve, left axis) and band gap (green curve, right axis) for $CaRuO_3$ (left plot) and $SrRuO_3$ (right plot) as a function of the $U_{eff}$ parameter from DFT +U computations (line are guide for the eyes). It shows that the transition from ferromagnetic to antiferromagnetic ground state occurs for $U_{eff} = 1.9eV$ ($U_{eff} = 2.6eV$) for $CaRuO_3$ ($SrRuO_3$), respectively. Concomitantly the system undergo a metal to insulator transition at $U_{eff}\sim1.9eV$ suggesting equivalent Ru-O covalence for both $CaRuO_3$ and $SrRuO_3$. Owing experimental results showing low band gap and dominance of AFM (FM) in $CaRuO_3$ ($SrRuO_3$), respectively, a reasonable $U_{eff}$ might be chosen between $2eV < U_{eff} < 2.5$ eV.

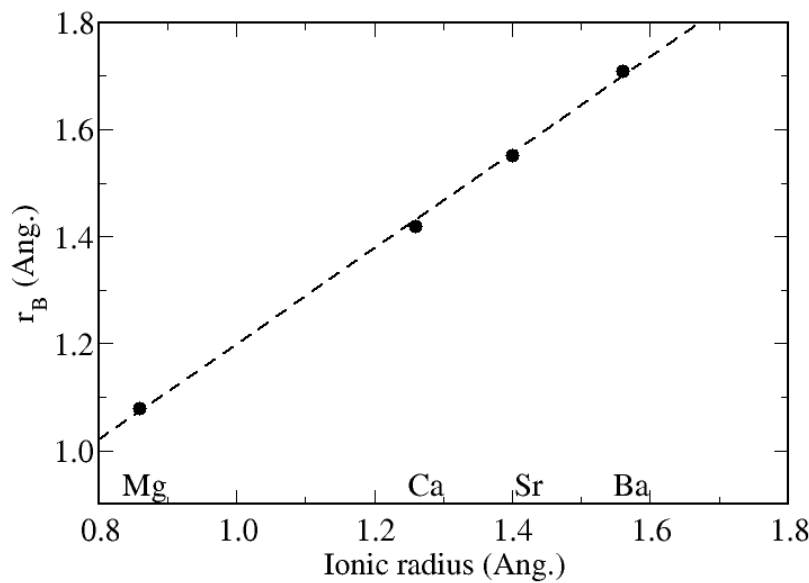

**Figure S6**: Bader radius of the alkali earth computed within DFT+U ($U_{eff} = 2.25eV$) as a function of ionic radius of the alkali-earth in perovskites $ARuO_3$ (A=Mg, Ca, Sr, and Ba). The dashed line shows the linear correlation between both radii.